\documentclass[useAMS,usenatbib]{mn2e}
\usepackage{graphicx,amssym}
\newcommand{\bc}{\begin{center}}
\newcommand{\ec}{\end{center}}

\title[The fraction of ellipticals in clusters]
      {What determines the fraction of elliptical galaxies in clusters?} 
\author[G.~De Lucia et al.]
       {Gabriella De Lucia$^{1}$\thanks{Email: delucia@oats.inaf.it}, 
        Fabio Fontanot$^{1}$ and David Wilman$^{2}$
        \\  
        $^1$INAF - Astronomical Observatory of Trieste, via G.B. Tiepolo 11, 
        I-34143 Trieste, Italy\\
        $^2$Max-Planck-Institut f\"ur Extraterrestrische Physik, 
        Giessenbachstra\ss e, D-85748 Garching, Germany} 
\begin{document}

\pagerange{\pageref{firstpage}--\pageref{lastpage}} 
\pubyear{2011}

\maketitle

\label{firstpage}

\begin{abstract}
We study the correlation between the morphological mix of cluster galaxies and
the assembly history of the parent cluster by taking advantage of two
independently developed semi-analytic models for galaxy formation and
evolution. In our models, both the number of cluster members and that of
elliptical members increase as a function of cluster mass, in such a way that
the resulting elliptical fractions are approximately independent of cluster
mass. The population of cluster ellipticals exhibit a marked bimodal
distribution as a function of galaxy stellar mass, with a dip at masses $\sim
10^{10}\,{\rm M}_{\odot}$. In the framework of our models, this bimodality
originates from the combination of a strongly decreasing number of galaxies
with increasing stellar mass, and a correspondingly increasing probability of
experiencing major mergers. We show that the correlation between the measured
elliptical fraction and the assembly history of the parent cluster is weak, and
that it becomes stronger in models that adopt longer galaxy merger times. We
argue that this results from the combined effect of a decreasing bulge
production due to a reduced number of mergers, and an increasing survival
probability of pre-existing ellipticals, with the latter process being more
important than the former.
\end{abstract}


\section{Introduction}
\label{sec:intro}

It has long been known that early type galaxies (ellipticals and lenticulars)
reside preferentially in dense regions of the Universe such as rich clusters,
while late type galaxies represent a larger fraction of the galaxy population
inhabiting regions of `average' density. Such a morphology-density relation was
noticed in early observational studies \citep[indications of a correlation
  between the type of {\it nebulae} and the environment can be found in `The
  Realm of Nebulae' by][]{Hubble_1936}, and was firmly established by
\citet{Dressler_1980}. 

In the past decades, much observational information has been collected on the
morphological distributions of cosmic galaxy populations, and on its dependence
on the environment. \citet{Butcher_and_Oemler_1978a,Butcher_and_Oemler_1984}
showed, for the first time, that the fraction of blue (star forming) galaxies
in clusters increases with increasing redshift. Detailed morphological studies
have been carried out in the following years, demonstrating that the fraction
of spiral galaxies increases with increasing redshift, and that this increase
appears approximately balanced by a decrease in the fraction of the lenticular
galaxies since $z\sim 0.5$. Over the same redshift range, the fraction of
elliptical galaxies is approximately constant
\citep{Dressler_1997,Fasano_etal_2000}. In more recent years, detailed
morphological studies have been pushed to lower mass ranges
\citep{Wilman_etal_2009}, and to higher redshift, where the mean fraction of
different morphological types does not appear to evolve significantly
\citep{Postman_etal_2005,Desai_etal_2007}.

At a given redshift, clusters with similar mass (measured from either X-ray
luminosity, or velocity dispersion) exhibit a non negligible scatter in their
morphological composition \citep[e.g.][]{Poggianti_etal_2009}. In the context
of the currently accepted paradigm for structure formation (the $\Lambda$CDM
model), it is logical to relate this cluster-to-cluster variance to the
dynamical history of the cluster. Although difficult to test quantitatively,
this expectation is confirmed by early observations that centrally concentrated
clusters have typically large populations of ellipticals and lenticulars and
relatively low numbers of spirals, while irregular, unrelaxed clusters are more
spiral-rich and show weaker radial gradients in their morphological mix
\citep[e.g.][]{Butcher_and_Oemler_1978b}. In this paper, we will address this
issue by considering two different semi-analytic models of galaxy formation,
and by relating the predicted fraction of elliptical galaxies to the accretion
history of the simulated cluster haloes.

\section{The galaxy formation models}
\label{sec:simsam}

In this paper, we take advantage of two independently developed galaxy
formation models: the `Munich' model, with the implementation discussed in
\citet{DeLucia_and_Blaizot_2007} and applied to the Millennium Simulation, and
the {\sc MORGANA} model, as adapted to the WMAP3 cosmology in
\citet{LoFaro_etal_2009}. Hereafter, we will refer to the former model as
DLB07. We note that in previous work, we have used the models presented in
\citet{Wang_etal_2008} which corresponds to the model by
\citet{DeLucia_and_Blaizot_2007} used here, but has been adapted to a WMAP3
cosmology. In this paper, we use the model applied to the Millennium Simulation
as this provides a larger volume and therefore a larger number of massive
haloes.

The simulations employed in this study assume a different cosmology: the
Millennium Simulation assumed a cosmological model that is consistent with WMAP
first-year results\footnote{The most important difference between WMAP first
  and third year data is a lower value for the amplitude of matter fluctuations
  on $8\,h^{-1}\,{\rm Mpc}$ scale ($\sigma_8$), which leads to a delay in
  structure formation \citep[e.g.][]{Wang_etal_2008}.}. As shown in previous
work, however, once the model is re-tuned to account for the change in
cosmology, the basic results and trends do not change significantly
\citep{Wang_etal_2008}. In addition, we note that the two models adopt
different definitions for the halo mass: in DLB07, this is given by $M_{200}$
and is computed from the simulation outputs as the mass contained in a sphere
of radius $R_{200}$, for which the mean overdensity is $200$ times the critical
density of the Universe at the redshift of interest. For {\sc MORGANA}, the
masses are simply given by the sum of the particle mass associated with the
halo, computed using {\small PINOCCHIO} \citep{Monaco_etal_2002}.

In this section, we provide a brief summary of the model elements that are
relevant to the present study. We refer to the original papers for a more
detailed discussion of the physical processes considered, and of the
corresponding modelling adopted. Both models consider two different channels
for the formation of bulges: galaxy-galaxy mergers and disk instabilities. The
relative importance of these channels, in different environments and at
different times, has been studied in detail in \citet{DeLucia_etal_2011}, while
in \citet{Fontanot_etal_2011} we focused on the statistics and properties of
{\it bulgeless} galaxies. Both models used in this study assume a Chabrier
Initial Mass Function.

Mergers are classified as {\it minor} or {\it major} according to their
baryonic (gas + stars) mass ratio. If this is smaller than 0.3, the merger is
classified as minor: the stellar mass of the secondary is added to the bulge
component of the primary galaxy, and the merger is accompanied by a
starburst. The resulting stars are added to the bulge component (in {\sc
  MORGANA}) or to the disk component (in DLB07). If the baryonic mass ratio of
the merging galaxies is larger than $0.3$, we assume that we witness a major
merger. In this case, both models assume that the disk components of the
merging galaxies are completely destroyed. The remnant spheroidal galaxy can
re-grow a new disk, if fed by an appreciable cooling flow. In previous work, we
have found that the merger model adopted in {\sc MORGANA} provides merging
times that are systematically shorter (by up to an order of magnitude) than
those adopted in the DLB07 model (see Section 7 of
\citealt{DeLucia_etal_2010}). The shorter merger times adopted in {\sc MORGANA}
lead to a more efficient formation of bulges and to larger number densities of
early type galaxies, particularly at high redshift. As we will show below, the
different modelling adopted for galaxy mergers also affects the relation
between the morphological fraction and the accretion history of dark matter
haloes. In order to quantify the significance of this effect, in the following
we will also show or discuss the results obtained from {\sc MORGANA} using the
same dynamical friction timescale prescription adopted by DLB07.

The treatment of disk instability differs significantly in the two models
considered: both adopt the same stability criterion proposed in
\citet*{Efstathiou_Lake_Negroponte_1982} but use different definitions for the
relevant physical quantities, and make different assumptions about the outcome
of instabilities: DLB07 only transfer to the bulge a fraction of the stellar
disc that is enough to restore stability. In the {\sc MORGANA} model, half of
the disk baryonic mass (both gas and stars) is transferred to the bulge. As
discussed and shown in \citet{DeLucia_etal_2011}, this translates into a more
relevant contribution of disk instability to bulge formation.

In the framework of our models, most of the elliptical galaxies acquire their
morphology through major mergers. Disk instability can contribute significantly
for low and intermediate mass galaxies, depending on the adopted treatment for
galaxy mergers and instabilities. As mentioned above, bulge dominated galaxies
can later grow a new disc, if they are fed by an appreciable cooling flow. We
have shown that the rates of disc regrowth are negligible for massive galaxies
and at low redshift. They represent, however, a non-negligible component of the
evolution of low and intermediate mass galaxies, particularly at high redshift
(see Section 6 of \citealt{DeLucia_etal_2011}). As we focus on galaxy clusters,
the model ellipticals considered in this paper are almost all satellite
galaxies (with the exclusion of central cluster galaxies). For these galaxies,
the bulge-to-total ratio is not affected after accretion onto a more massive
halo in the {\sc MORGANA} model. DLB07 accounts for mergers between satellites
(that are, however, rare) so that the bulges of satellite galaxies can still
grow through this physical mechanism. Finally, none of the models used in this
study include `environmental' processes such as tidal stripping or harassment,
that can potentially affect the morphology of satellite galaxies orbiting in a
massive cluster \citep[e.g.][]{Mastropietro_etal_2005}.

In our previous work, we have considered alternative prescriptions to model
bulge formation, including predictions obtained when the disk instability
channel is switched off. We have verified that the results presented in the
following do not depend significantly on these assumptions. Therefore, we will
discuss only results obtained by our default models. As these data have not
been used to `tune' the models in the first place, they can be considered as
genuine model predictions, and compared with available observational
measurements.

\section{The fraction of elliptical galaxies in clusters}
\label{sec:efrac}

\begin{figure*}
\bc
\resizebox{18cm}{!}{\includegraphics[]{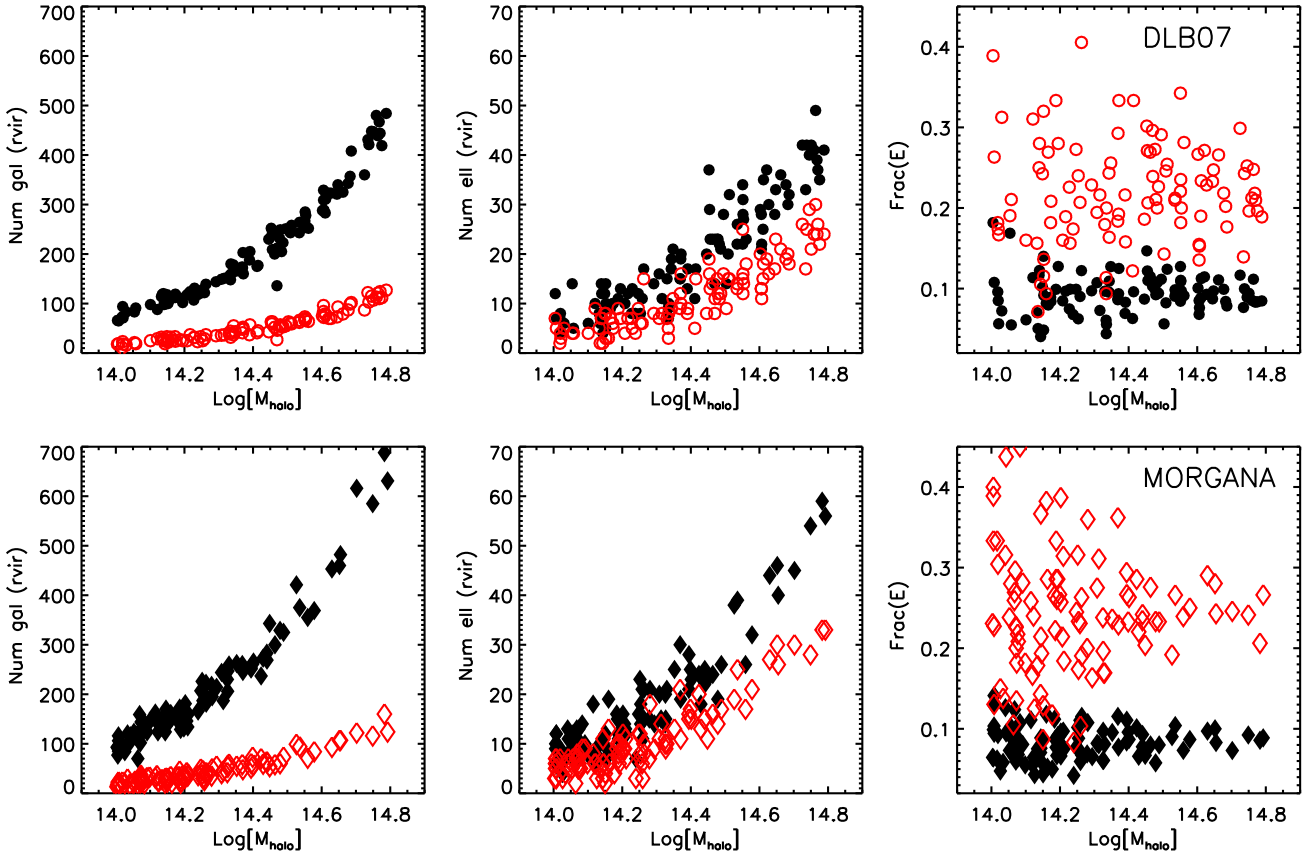}} 
\caption{Total numbers of galaxies (left panel), numbers of ellipticals (middle
  panel), and elliptical fractions (right panel) inside the cluster virial
  radius, as a function of the cluster mass. Filled and open symbols (these are
  coloured black and red in the online edition of the Journal) are obtained
  considering all galaxies more massive than $10^9$ and $10^{10}\, {\rm
    M}_{\odot}$, respectively.}
\label{fig:efrac}
\ec
\end{figure*}

In this study we have considered galaxy clusters with masses in the range ${\rm
  log \, M/M_{\odot}}= 14 - 14.8$, at redshift zero. In the simulation used by
{\sc MORGANA}, there are $100$ clusters over this mass range. To compare the
predictions from this model to those from DLB07, we have selected the same
number of haloes from the Millennium Simulation, uniformly distributed in mass
over the same mass range. In the following, we will define as {\it ellipticals}
all galaxies that have a stellar bulge to total ratio larger than $0.9$. When
relevant, we will comment on how results can be affected by a different
threshold.

Figure~\ref{fig:efrac} shows the total number of galaxies (left panel), the
number of ellipticals (middle panel), and the elliptical fractions (right
panels) as a function of the cluster virial mass. Only galaxies residing within
$R_{200}$ have been considered in the DLB07 model. Since {\sc MORGANA} does not
provide information on the position of galaxies within dark matter haloes, we
have simply considered in this case all galaxies associated with the final
cluster. Given the different definitions adopted, any difference between model
predictions (but we will see that these are very small) should be interpreted
with caution. Filled and open symbols in Figure~\ref{fig:efrac} correspond to
galaxies more massive than $10^9$ and $10^{10}\, {\rm M}_{\odot}$,
respectively. The former limit corresponds to the approximate resolution limit
of the Millennium Simulation, while the latter corresponds to a typical limit
for observational studies.

Figure~\ref{fig:efrac} shows that both the total number of galaxies and the
number of ellipticals increase with increasing halo mass. When a lower stellar
mass threshold is chosen, the predicted numbers are significantly higher. A
cluster of mass $2.5\times10^{14}\,{\rm M}_{\odot}$ contains on average $\sim
50$ ($\sim 60$) galaxies more massive than $10^{10}\, {\rm M}_{\odot}$ in the
DLB07 ({\sc MORGANA}) model. When considering a mass limit of $10^{9}\, {\rm
  M}_{\odot}$, the average number of cluster members within the virial radius
increases to $\sim 190$ ($\sim 260$ in {\sc MORGANA}). The number of
ellipticals increases too, but not as much as the total number of
galaxies. This is expected given that, as the stellar mass increases, a larger
fraction of galaxies are classified as ellipticals (see Fig. 7 in
\citealt{DeLucia_etal_2011}). Interestingly, the halo occupation distribution
of the two models used in this study is different, with {\sc MORGANA} always
predicting a slightly larger number of cluster members with respect to
DLB07. The difference is significant for the most massive clusters included in
our sample, and when considering all galaxies more massive than $10^{9}\, {\rm
  M}_{\odot}$. We stress, however, that a different definition for cluster
members has been adopted for the two models. In addition, we are using the
dynamical information available from the simulations to define cluster members,
while an accurate comparison with observational measurements should account for
possible contamination by interlopers along the line of sight.  As mentioned
above, the merger times adopted in {\sc MORGANA} are about one order of
magnitude shorter than those adopted in DLB07. By using longer merger times in
{\sc MORGANA}, the number of cluster members increases even further.

The fraction of elliptical galaxies resulting from the numbers shown in the
left and middle panel of Figure~\ref{fig:efrac} does not vary significantly as
a function of cluster mass, in agreement with observational measurements in the
local Universe \citep{Wilman_etal_2009,Poggianti_etal_2009}\footnote{We note
  that both \citet{Wilman_etal_2009} and \citet{Poggianti_etal_2009} are based
  on magnitude limited samples, while we are using a cut on total stellar
  mass.}. The predicted elliptical fractions are of the order of $10$ per cent
in both models, when all galaxies more massive than $10^{9}\, {\rm M}_{\odot}$
are considered. For a mass threshold of $10^{10}\, {\rm M}_{\odot}$, the
expected fractions increase, and the scatter becomes larger. Interestingly, the
halo to halo scatter appears to increase slightly with decreasing halo
mass. This is more evident in the {\sc MORGANA} model, but we note that in this
case haloes are not distributed uniformly in mass and the number of clusters at
the largest masses considered is quite low. Therefore, the very narrow range of
elliptical fractions predicted by this model for the most massive haloes might
be fortuitous, and just due to poor number statistics.

When considering all galaxies more massive than $10^{10}\, {\rm M}_{\odot}$,
the mean elliptical fraction is $0.22$ for the DLB07 model, and $0.24$ for {\sc
  MORGANA}. This is lower than the average value of $\sim 0.32$ measured for
the WIde-field Nearby Galaxy clusters Survey (WINGS), using a similar mass cut
\citep{Vulcani_etal_2011}. We note, however, that only galaxies within
$0.6\,{\rm R}_{200}$ have been considered in this observational study, as this
is the largest radius covered in all their cluster fields, and that the study
is based on a definition of ellipticals that differs from that adopted in this
paper (morphologies have been assigned using V-band images). In previous work
\citep{Simard_etal_2009}, we have shown that the early-type fractions predicted
by the DLB07 model compare well to observational measurements from the Sloan
Digital Sky Survey (SDSS) in the local Universe and from the ESO Distant
Cluster Survey (EDisCS) at redshift $\sim 0.6$. Also in that study, however, a
different (closer to that used in the observations) definition of `early-type'
galaxies was adopted, so that the predicted fractions shown in this study are
not the same as those shown in Simard et al. In a forthcoming paper (Wilman et
al., in preparation), we will carry out a more detailed comparison between the
observed mix of different morphological classes and predictions from our galaxy
formation models.

\begin{figure*}
\bc
\resizebox{16cm}{!}{\includegraphics[]{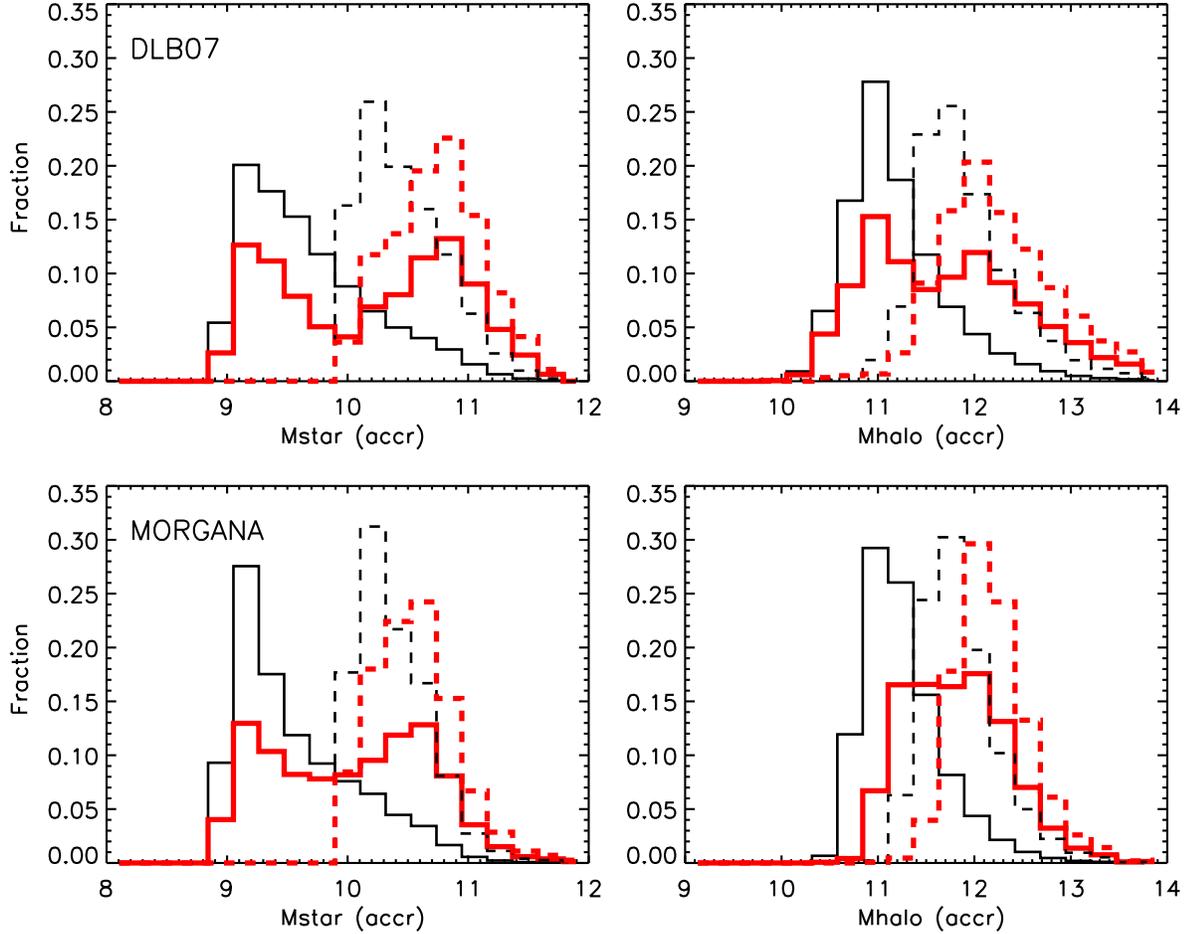}} 
\caption{Distribution present day stellar masses (left panel) for model
  ellipticals in our cluster sample, and for their parent halo mass at the time
  of accretion (right panels). The distributions shown have been obtained by
  stacking all clusters in the sample, and normalizing to the total number in
  each distribution. Solid and dashed lines refer to the cases when all
  galaxies more massive than $10^9$ and $10^{10}\, {\rm M}_{\odot}$ are
  considered, respectively. Thin and thick lines (black and red in the online
  edition of the Journal) are for all galaxies and for those that are
  classified as ellipticals, respectively.}
\label{fig:accrdistr}
\ec
\end{figure*}

The left panels of Figure~\ref{fig:accrdistr} show the predicted distributions
of stellar masses for all cluster galaxies (thin histograms) and for the
cluster ellipticals (thick histograms).  These distributions have been obtained
by stacking the galaxies in all clusters, and have been normalized to the total
number of galaxies in each distribution.  The two models provide very similar
predictions, but those from {\sc MORGANA} are more skewed towards less massive
galaxies. Interestingly, both models predict a bimodal distribution for
elliptical galaxies, with a pronounced `dip' around $\sim 10^{10}\, {\rm
  M}_{\odot}$. Unfortunately, this is below or approximately at the limit of
the observational measurements for the WINGS sample used in
\citet{Vulcani_etal_2011}. We note that this bimodal behaviour is found in our
models also when considering the global elliptical population (i.e. not only
ellipticals in clusters), which does not appear to be supported by available
observations \citep[e.g.][]{Trentham_and_Hodgkin_2002,Driver_etal_2003}. We
stress that our models (like most of the recently published models) overpredict
the number densities of small to intermediate mass galaxies
\citep{Fontanot_etal_2009}, so the importance of the peak at small masses is
likely over-estimated.

In both models, the bimodal distribution of the cluster elliptical masses is
significantly reduced (but still apparent in the {\sc MORGANA} model) when the
adopted threshold for defining a bulge dominated galaxy as an elliptical is
lowered to $\sim 0.7$. In this case, only one peak is visible in the DLB07
predictions at masses ${\rm log}(M_{\rm star}) \sim 10.5$. We have verified
that, in our model, this bimodality is not significantly affected when the disk
instability channel for bulge formation is switched off, so a differential
efficiency of bulge formation through disk instability is not responsible for
the shape of the cluster elliptical mass distribution shown in
Figure~\ref{fig:accrdistr}. In our previous work \citep{DeLucia_etal_2011}, we
have shown that disk regrowth is more efficient for intermediate mass
galaxies. In order to test if this could be responsible for the observed dip at
intermediate masses, we have calculated the mass distribution of all galaxies
that have been ellipticals in their past, either considering only those
surviving at redshift zero (i.e. excluding those that have merged with other
galaxies) or all galaxies in the merger trees of the cluster ellipticals. In
both cases, we find that the predicted mass distribution exhibit a marked
bimodality, with a pronounced `deficit' of elliptical galaxies at intermediate
masses.

We interpret this bimodality as a result of the increasing probability of
suffering a major merger with increasing mass (see Figure 9 in
\citealt{DeLucia_etal_2006} and Figure 6 in \citealt{Wang_and_Kauffmann_2008}),
and of the strongly decreasing number of galaxies of larger masses (as shown by
the thin lines in Figure~\ref{fig:accrdistr}). The convolution of these
distribution functions results in a lower number of intermediate mass galaxies
suffering of major merger events in their past history, compared to galaxies
residing in the low and high mass peaks of the distributions shown in the left
panels of Figure~\ref{fig:accrdistr}. In particular, Figure 3 of
\citet{Wang_and_Kauffmann_2008} shows that the region where the dip in the mass
distribution of elliptical galaxies is visible, corresponds to a regime where
the probability of suffering of a minor merger is significantly larger than
that of experiencing a major merger. This happens because, during the time that
elapses between a halo merger and the actual merger between the galaxies
residing at the centre of the merging haloes, the stellar mass of the satellite
does not increase significantly, while the central galaxy grows in mass as it
is fed by cooling from the surrounding hot halo. As a consequence, the stellar
mass ratio between the two galaxies decreases, so that a major merger between
two haloes can lead to a minor merger between their galaxies.  Figure 3 in
\citet{Wang_and_Kauffmann_2008} shows that the probability of experiencing a
minor merger is largest at intermediate masses, which explains why lowering the
adopted bulge-to-total threshold fills the intermediate region, and tends to
wash out the bimodality.

For each cluster member in our sample, we have traced back their main
progenitor until the galaxy is for the last time a central galaxy of a dark
matter halo, and we have recorded the parent halo mass at this time. In the
following, we refer to this as the `time of accretion', although this will not
always coincide with the time when the galaxy is accreted onto the main
progenitor of the final cluster (De Lucia et al., in preparation). The right
panels of Figure~\ref{fig:accrdistr} show the distributions of halo masses at
accretion for all cluster members (thin lines) and for the ellipticals (thick
lines). Again, the distributions obtained for all clusters have been stacked
and normalized to the total number of galaxies in each of them. 

The figure shows that, in both models, elliptical galaxies tend to be accreted
when they reside in more massive haloes with respect to the total population of
cluster members (the distribution predicted by {\sc MORGANA} has a higher lower
mass limit than in the DLB07 model: i.e. in {\sc MORGANA}, ellipticals tend to
be accreted, on average, in more massive haloes than in the DLB07 model).
This is not surprising considering that ellipticals represent a larger fraction
of the most massive galaxies, and that there is a relatively tight correlation
between the galaxy mass and that of the parent halo mass for central galaxies
\citep[e.g.][]{Wang_etal_2006}. 

\section{The assembly of clusters with different elliptical fractions}

In the previous section, we have shown that ellipticals tend to be accreted in
larger haloes, with respect to the entire cluster galaxy population. Given
these results, one would expect naively that haloes that have acquired a larger
fraction of their mass through the accretion of `massive' haloes would host a
larger fraction of ellipticals. One has to consider, however, that elliptical
galaxies can also {\it disappear} from the sample of cluster members by merging
with the central galaxies of the hosting halo (or with other satellites in the
DLB07 model).

\begin{figure*}
\bc
\resizebox{18cm}{!}{\includegraphics[]{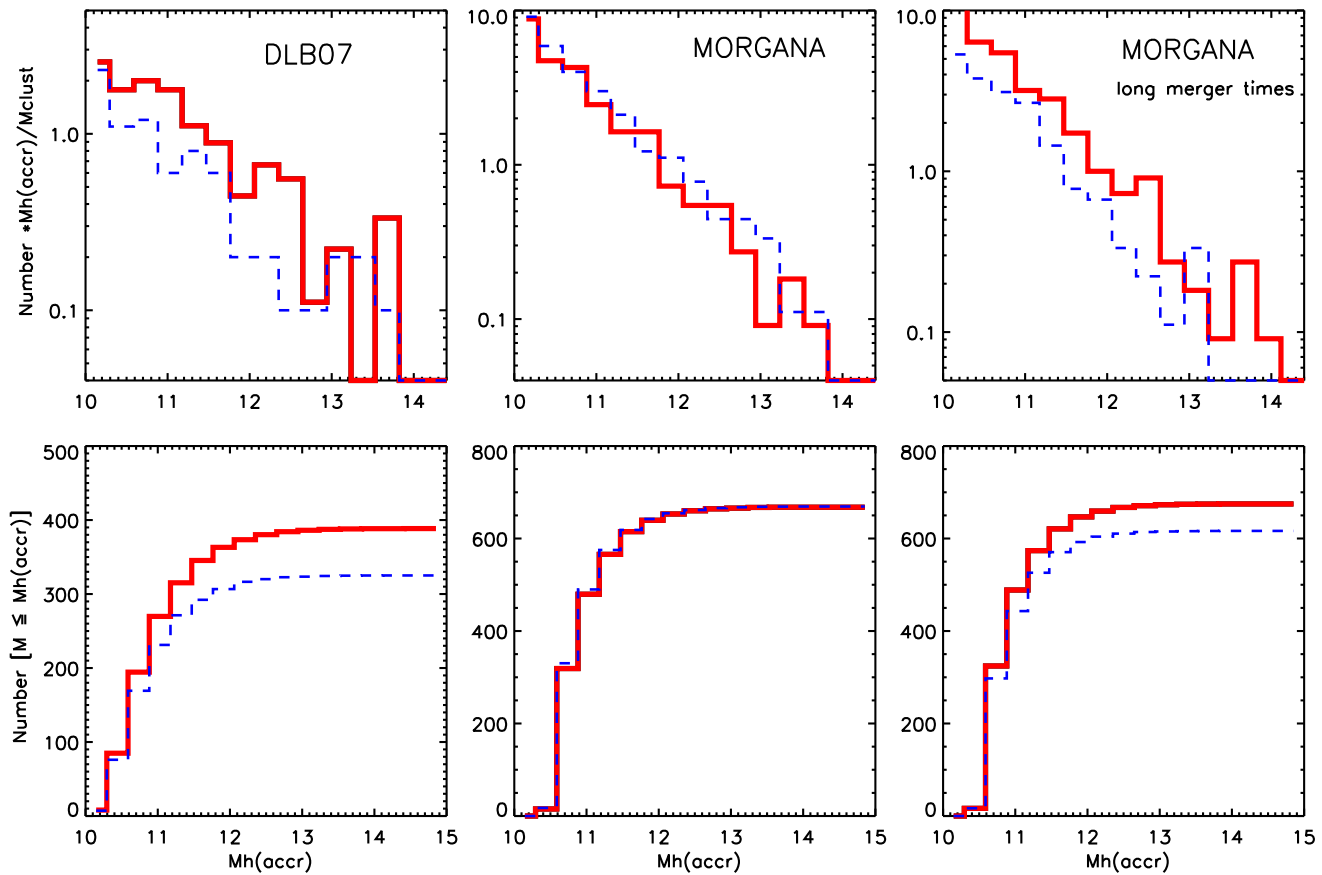}} 
\caption{Mass distribution of haloes accreted on the main branch of each
  cluster in our sample. Thin dashed lines (blue in the online edition of the
  Journal) are for haloes with an elliptical fraction that is lower than the
  10th percentile of the distribution, while thick solid lines (red in the
  online edition of the Journal) correspond to the haloes whose elliptical
  fraction is larger than the 90th percentile of the distribution. The
  differential distributions in the top panels have been weighted by mass, in
  order to remove the dominant mass dependence and emphasize the differences
  between the two samples.}
\label{fig:haloaccrdistr}
\ec
\end{figure*}

In order to address this issue, we have analysed the accretion histories of all
clusters included in our sample. For each halo, we have traced back in time its
main branch, i.e. the branch of the tree that is obtained by connecting the
halo to its {\it main progenitor}\footnote{In {\sc MORGANA}, this is simply the
  most massive progenitor at each node of the tree. A different definition is
  adopted in DLB07: the main progenitor is selected by choosing the branch of
  the merger tree that accounts for most of the mass of the final system, for
  the longer period. As explained in \citet{DeLucia_and_Blaizot_2007}, this
  avoids possible problems arising when the selection of the most massive
  progenitor would be ambiguous like, for example, when there are two
  progenitors of similar mass.}. We have then considered all substructures
residing in the main branch at each time, and have traced each of them back in
time until they were {\it main haloes} of a FOF-group. The top panels of
Figure~\ref{fig:haloaccrdistr} show the mass distributions of accreted haloes
for the clusters that host an elliptical fraction lower (thin dashed lines)
than the 10th percentile, and higher (thick solid lines) than the 90th
percentile of the distribution of elliptical fractions measured for all 100
haloes considered. Again, the distributions from the two cluster samples have
been stacked. Different columns correspond to different models, as indicated by
the legend, while the bottom panels show the corresponding cumulative
distributions. The differential distributions shown in the top panels of
Fig.~\ref{fig:haloaccrdistr} have been weighted by mass in order to remove the
dominant mass dependency and emphasize the differences between the two samples.

In the DLB07 model (left panels), there is a clear difference between the two
samples, which is more evident when looking at the cumulative distributions:
clusters that host the highest elliptical fractions also accreted a larger
number of haloes more massive than $\sim 10^{11}\,{\rm M}_{\odot}$, with
respect to the clusters that host the lowest elliptical fractions.
Interestingly, the difference between the two samples persist over the entire
mass range: this implies a lower contribution from diffuse accretion for
clusters with large elliptical fractions. In the standard {\sc MORGANA} model,
no significant difference is found between the two samples. If, however, longer
merger times are adopted (right panels), then a difference between the haloes
with largest and lowest elliptical fractions become visible, and it is of the
same order of magnitude of that found in the DLB07 model. At first sight, this
result appears counter-intuitive because one would expect that shorter merger
times would translate into a better matching between the morphological mix of
the galaxy population and the assembly history of the halo. One has to
consider, however, that changing the merger times would affect the elliptical
galaxy population in two distinct ways: on the one hand, longer merger times
would tend to decrease the number of mergers (and therefore the number of bulge
dominated galaxies). On the other hand, longer merger times would also tend to
`preserve' the pre-existing ellipticals from being accreted onto the central
galaxies (or from merging with other satellites if this physical process is
included). When adopting longer merger times in {\sc MORGANA}, we find that the
second process would be slightly more important than the first. As a
consequence, both the total number of cluster members and the number of
elliptical members would increase. This implies that longer merger times
preserve a better memory of the accretion history of the parent dark matter
haloes, thereby creating a stronger correlation between the morphological mix
of the cluster galaxy population and its dynamical status.

\section{Discussion and conclusions}
\label{sec:discconcl}

At fixed cluster mass, the observed properties of the cluster galaxy population
exhibit a large variation. Such a scatter is in part due to observational
uncertainties in the observed quantities. In the hierarchical framework,
however, it is natural to link the observed halo-to-halo scatter to a range of
dynamical histories of the parent cluster. In this paper, we have investigated
the link between the predicted fraction of elliptical galaxies and the
accretion history of the parent dark matter halo, by taking advantage of two
different semi-analytic models of galaxy formation. Our main results can
  be summarized as follows:

\begin{itemize}

\item For both models used in our study, the predicted elliptical fractions do
  not vary significantly as a function of the cluster mass, for the range of
  masses considered ($M_{200} \gtrsim 10^{14}\,{\rm M}_{\odot}$). This appears
  to be in qualitative agreement with observational measurements
  \citep{Wilman_etal_2009,Poggianti_etal_2009,Simard_etal_2009}. In our models,
  a constant elliptical fraction results from an increasing number of both
  cluster members and elliptical members, as a function of cluster mass.

\item Cluster ellipticals exhibit a marked bimodal distribution in stellar
  mass. In both models, the bimodality is reduced when a lower ($\sim 0.7$)
  bulge-to-total threshold is adopted for selecting elliptical galaxies. The
  distribution of stellar masses for elliptical galaxies preserves its bimodal
  behaviour when considering all galaxies (i.e. is not limited to cluster
  ellipticals).

\item Since ellipticals are the dominant population among massive cluster
  members, one finds that these galaxies have been accreted, on average, onto
  the cluster when residing in relatively massive structures (more massive than
  those of the overall cluster galaxy population). This creates a correlation
  between the observed fraction of ellipticals and the accretion history of the
  halo that is, however, not strong.
\end{itemize}

In the framework of our models, the bimodal distribution of elliptical stellar
masses is not due to a more prominent role played by disk instability and/or
disk regrowth for intermediate mass galaxies \citep{DeLucia_etal_2011}. We
argue that this bimodality results simply from the convolution between the
strongly decreasing number of galaxies and the increasing probability of
experiencing a major merger event in the past, for increasing galaxy mass
\citep{DeLucia_etal_2006,Wang_and_Kauffmann_2008}. For galaxies with mass $\sim
10^9\,{\rm M}_{\odot}$, the probability of having experienced a major merger is
not large, but there are many low-mass galaxies. On the other hand, almost all
galaxies with mass $\gtrsim 10^{11}\,{\rm M}_{\odot}$ have experienced at least
one major merger during their lifetime so, although the number of massive
galaxies is low, this mass bin is dominated by elliptical galaxies.

We argue that clusters that host larger fraction of ellipticals have a lower
contribution from diffuse accretion than clusters with lower elliptical
fractions (i.e. they accrete more haloes, over the entire mass range probed by
our simulations). In addition, we find that the correlation between the
observed fraction of elliptical galaxies and the accretion history of the halo
can be weakened in the case of short merger times. This would reduce the number
of elliptical cluster members by having them accreted onto the central cluster
galaxies or merged with other cluster members. In this framework, elliptical
satellites have been formed {\it before} their accretion onto the cluster. The
measured fraction of ellipticals is determined by the balance between the
disappearance of ellipticals due to accretion and mergers, and the recent
accretion of relatively massive structures (that would likely host an
elliptical central galaxy). A better `memory' of the accretion history is
preserved when merger times are longer. In this case, a stronger correlation
between the morphological mix of cluster populations and the dynamical status
of the cluster is expected.

\section*{Acknowledgements}

The Millennium Simulation databases used in this paper and the web application
providing online access to them were constructed as part of the activities of
the German Astrophysical Virtual Observatory.  GDL acknowledges financial
support from the European Research Council under the European Community's
Seventh Framework Programme (FP7/2007-2013)/ERC grant agreement n. 202781. FF
acknowledges the support of an INAF-OATs fellowship granted on `Basic Research'
funds and financial contribution from the ASI project `IR spectroscopy of the
Highest Redshift BH candidates' (agreement ASI-INAF 1/009/10/00).  DW
acknowledges the support of the Max-Planck Gesellschaft. We thank Pierluigi
Monaco and Simone Weinmann for useful discussions.

This paper is dedicated to my grandmother.

\bsp

\label{lastpage}

\bibliographystyle{mn2e}
\bibliography{erich_epoor}

\end{document}